\documentclass[preprint,12pt]{elsarticle}

\usepackage{amsmath,amssymb,amsfonts}
\usepackage{graphicx}

\journal{Optics Communications}

\begin{document}

\begin{frontmatter}

\title{A photonic crystal realization of a phase driven two-level atom}

\author{B. M. Rodr\'{\i}guez-Lara}
\ead{bmlara@inaoep.mx}
\author{Alejandro Z\'{a}rate C\'{a}rdenas, Francisco Soto-Eguibar, H\'{e}ctor Manuel Moya-Cessa}
\address{Instituto Nacional de Astrof\'{\i}sica, \'Optica y Electr\'onica, Calle Luis Enrique Erro No. 1, Sta. Ma. Tonantzintla, Pue. CP 72840, M\'exico}

\begin{abstract}
We propose a set of photonic crystals that realize a nonlinear quantum Rabi model equivalent to a two-level system driven by the phase of a quantized electromagnetic field.
The crystals are exactly solvable in the weak-coupling regime; their dispersion relation is discrete and the system is diagonalized by normal modes similar to a dressed state basis. 
In the strong-coupling regime, we use perturbation theory and find that the dispersion relation is continuous. 
We give the normal modes of the crystal in terms of continued fractions that are valid for any given parameter set.
We show that these photonic crystals allow state reconstruction in the form of coherent oscillations in the weak-coupling regime.
In the strong-coupling regime, the general case allows at most partial reconstruction of single waveguide input states, and non-symmetric coherent oscillations that show partial state reconstruction of particular phase-controlled states. 
\end{abstract}


\end{frontmatter}

\section{Introduction}

Photonic crystals as classical simulators of quantum processes have been the focus of attention in recent years~\cite{Longhi2006p110402,Perets2008p170506,Bromberg2009p253904,Dreisow2009p076802,Lahini2010p163905,Dreisow2010p143902,Longhi2010p075102,Longhi2011p3248,Keil2011p103601,RodriguezLara2011p053845,Longhi2011p453,Longhi2012p435601,Longhi2012p012112,Garanovich2012}.
In particular, it has been shown theoretically and experimentally that the so-called quantum Rabi model describing the interaction of a two-level system with a quantum field may be realized by photonic superlattices~\cite{Longhi2011,Crespi2012p163601}.
The quantum Rabi model in the weak-coupling regime, i.e. the Jaynes-Cummings model~\cite{Jaynes1963p89}, describes a variety of quantum mechanical systems that have been experimentally implemented; e.g. cavity-quantum electrodynamics (cavity-QED)~\cite{Walther2006p1325}, ion traps~\cite{MoyaCessa2012p229} and  circuit-QED~\cite{Blais2004p062320}.
Strong-coupling is not feasible in a majority of simple quantum optical systems but photonic crystals provide a classical realization of the quantum model for all coupling regimes~\cite{Longhi2011,Crespi2012p163601}.

In quantum optics, diverse non-linear models describing the interaction between a two-level system and a quantum field have been proposed as deformations of the Jaynes-Cummings model~\cite{Kundu2004p281,delosSantosSanchez2012p015502}.
One example of these nonlinear models is the Buck-Sukumar (BS) model where the atom--field coupling depends on the intensity of the quantum field~\cite{Buck1981p132}.
The BS model, which is exactly solvable and does not have a feasible experimental representation, unless it is classically realized in a couple of binary photonic crystals where the coupling depends linearly on the position of the waveguide, helps in understanding the apparition of collapses and revivals of the two-level inversion in the radiation--matter interaction systems.

In the following, we propose a semi-infinite photonic crystal that classically simulates a novel non-linear quantum optics model describing an atom driven by just the phase of a quantum field. 
Up to our knowledge both the non-linear radiation-matter interaction model and its photonic realization are missing in the literature.
Then, we find the exact dispersion curves and normal modes of the waveguide lattice in the weak-coupling regime.
In the strong-coupling regime, the dispersion relation is continuous and we find the normal modes as continued fractions.
The transition from discrete to continuous spectrum, appearing in our photonic crystal, does not show in the spectra of Rabi~\cite{Tur2000p574,Tur2001p899,Casanova2010p263603} and BS~\cite{Buck1981p132} models which are discrete in both regimes, weak and strong coupling.
Thus, parameter sets delivering coherent oscillations in Rabi or BS models only produce coherent oscillations in the weak-coupling regime of our model.

%
\section{The model and its photonic crystal analogue}
Let us consider the Hamiltonian describing a two-level system driven by just the phase of a quantum field,
\begin{eqnarray}
\hat{H} = \omega_{f} \hat{a}^{\dagger} \hat{a} + \frac{\omega_{0}}{2} \hat{\sigma_{z}} + \lambda \left( e^{\imath \hat{\phi}} + e^{-\imath \hat{\phi}} \right) \hat{\sigma}_{x}, \label{eq:Hamiltonian}
\end{eqnarray}
where the exponential of the quantum phase operator is given by the Susskind-Glogower operator~\cite{Susskind1964p49}
\begin{eqnarray}
 e^{\imath \hat{\phi}} \equiv \hat{V} =  \frac{1}{\sqrt{\hat{a} \hat{a}^{\dagger}}} \hat{a}.
\end{eqnarray}
The field mode of frequency $\omega_{f}$ is described by the annihilation (creation) operators $\hat{a}$ ($\hat{a}^{\dagger}$), the two-level system of transition frequency $\omega_{0}$ by Pauli matrices $\sigma_{x,y,z}$, and their interaction by the real coupling $\lambda$.
It is possible to separate this system in two uncoupled Hamiltonians,
\begin{eqnarray}
\hat{H}_{\pm} = \omega_{f} \hat{n} \mp \frac{\omega_{0}}{2} (-1)^{\hat{n}} + \lambda \left( \hat{B} + \hat{B}^{\dagger} \right),
\end{eqnarray}
belonging to one of two parity chain basis,
\begin{eqnarray}
|+,n \rangle &=& \hat{B}^{\dagger n} | 0, g\rangle,  \\
|-,n \rangle &=& \hat{B}^{\dagger n} | 0, e\rangle,
\end{eqnarray}
defined such that parity,
\begin{eqnarray}
\hat{\Pi} = -\sigma_{z} (-1)^{\hat{n}},
\end{eqnarray}
is conserved, $\langle \pm ,n \vert \hat{\Pi} \vert \pm, n \rangle = \pm$; the bases annihilation (creation) operator is given by $\hat{B} = \hat{V} \hat{\sigma}_{x}$ ($\hat{B}^{\dagger} = \hat{V}^{\dagger} \hat{\sigma}_{x}$) and the number operator is defined as $\hat{n} \vert \pm, m \rangle = m \vert \pm, m \rangle $.
By defining the general state,
\begin{eqnarray}
\vert \psi_{\pm} \rangle = \sum_{j=0}^{\infty} \mathcal{E}_{j}^{(\pm)} \vert \pm, j \rangle,
\end{eqnarray}
the equations of motion for any given initial state under the dynamics given by Hamiltonian (\ref{eq:Hamiltonian}) are reduced to the differential set
\begin{eqnarray}
i \partial_{t} \mathcal{E}^{(\pm)}_{j} = \left[ \omega_{f} j \mp \frac{\omega_{0}}{2} (-1)^{j} \right] \mathcal{E}^{(\pm)}_{j} + \lambda \left( \mathcal{E}^{(\pm)}_{j-1} + \mathcal{E}^{(\pm)}_{j+1} \right), \label{eq:DiffSet}
\end{eqnarray}
where the shorthand notation $\partial_{t}$ has been used for the partial derivative with respect to $t$.
This differential set is equivalent, up to a phase and substituting $t \rightarrow z$, to that describing the propagation equation of a classical field through a photonic waveguide lattice. In this equivalent photonic waveguide lattice, $\mathcal{E}_{j}$ is the amplitude of the field at the $j$th waveguide, the waveguides are homogeneously coupled, and the refraction indices grow proportional to $\omega_{f}$ and to their position on the lattice plus a position depending bias proportional to $\omega_{0}/2$.

\begin{figure}[h!]
\centering \includegraphics [scale=1] {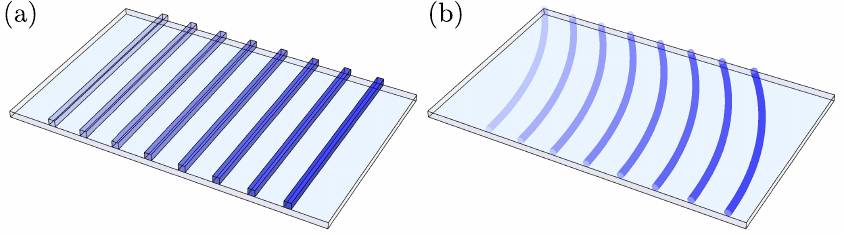}
\caption{(Color online) Two different schemes to produce the set of photonic crystals realizing the phase driven two-level atom. A semi-infinite set of homogeneously coupled waveguides where the refraction index behaves as the function $ n^{(\pm)}_{j} \propto \omega_{f} j \mp \frac{\omega_{0}}{2} (-1)^{j} $.  (a)  Straight waveguides deliver fixed $\omega_{f}$ and $\omega_{0}$ parameters. (b) Circularly bent waveguides produce a parameter $\omega_{f}$ that is proportional to the frequency of impinging light and a fixed parameter $\omega_0$.}
\label{fig:Fig1}
\end{figure}

For the sake of simplicity, hereby we will refer to the photonic crystals as $H_{+}$ or $H_{-}$, depending on the sign of Eq.(\ref{eq:DiffSet}).
In order to construct these photonic crystals, one can choose to either implement a static, Fig. \ref{fig:Fig1}(a), or dynamic, Fig. \ref{fig:Fig1}(b), relation between parameters $\omega_{f}$ and $\omega_{0}$. The parameter $\omega_{0}$ will be fixed in both cases. 
Straight waveguides produce a fixed parameter $\omega_{f}$ independent of the wavelength of the impinging light.
Bending the waveguides along a circle introduces an index gradient inversely proportional to the wavelength of the impinging light~\cite{Longhi2011,Crespi2012p163601}; thus, in this case the parameter $\omega_{f}$ depends on the wavelength of the impinging light and one can vary the ratio between $\omega_{f}$ and $\omega_{0}$ by choosing the color of the impinging light.

It is interesting to notice that in the case $\omega_{0}=0$ our differential set in Eq.(\ref{eq:DiffSet}) reduces to the one describing a semi-infinite waveguide lattice with linearly increasing refractive indexes that presents Bloch oscillations~\cite{Peschel1998p1701,Pertsch1999p4752}. 
It is also known that an infinite lattice, considering $\omega_{0}=0$ with a two-waveguides input with a phase difference between them, i.e. $\mathcal{E}(t=0) = \mathcal{E}_{0}(0) +  e^{i \phi} \mathcal{E}_{1}(0) $ with $\phi \neq 0$ and $\mathcal{E}_{0,1}(0) \in \mathbb{R}$, shows a ratchet-like behavior controlled by the phase $\phi$~\cite{Thompson2011p214302}.   
In our idea of classical simulation of a radiation-matter interaction system, it may be possible to argue that the case $\omega_{0}=0$ corresponds to emulating a hot trapped ion coupled to just the phase of a bosonic mode; such an argument has been used in the quantum Rabi problem~\cite{Casanova2010p263603}.

%
\section{Dispersion relation}

Up to our knowledge and means, it is not possible to find an exact dispersion relation for the photonic crystals described above, but it is possible to separate the relevant coupling parameter in two regimes, weak and strong, in order to obtain some results.
In the first of these regimes, we can borrow techniques from quantum optics and find an exact dispersion relation.
While on the last, we can only deal with the problem through perturbation theory.

%
\subsection{Weak-coupling regime: $\lambda \ll \omega_{f}, \omega_{0}$.}
In the weak coupling regime one can find the exact spectrum and normal modes of each one of the photonic crystals described by the differential set (\ref{eq:DiffSet}) by taking a step back and implementing the rotating wave approximation in the Hamiltonian of the system,
\begin{eqnarray}
\hat{H}_{RWA} = \omega_{f} \hat{a}^{\dagger} \hat{a} + \frac{\omega_{0}}{2} \hat{\sigma_{z}} + \lambda \left( e^{\imath \hat{\phi}} \hat{\sigma}_{+} + e^{-\imath \hat{\phi}} \hat{\sigma}_{-} \right), \label{eq:HRWA}
\end{eqnarray}
before establishing the classical analogue.
Then, it is simpler to find the spectrum and normal modes in this representation by using the basis set $\left\{ \vert n, e \rangle, \vert n+1, g \rangle \right\}$ belonging to the manifold with $n+1$ excitations.
This leads to the discrete spectrum,
 \begin{eqnarray}
E_{\pm,n} = \omega_{f} \left( n + \frac{1}{2} \right) \pm \frac{\Omega}{2}, \qquad \Omega = \sqrt{ \delta^2 + 4 \lambda^{2}}, \label{eq:EigValRWA}
\end{eqnarray}
where the detuning is defined as $\delta = \omega_{f} - \omega_{0}$. The proper states are given by
\begin{eqnarray}
\vert n, \pm \rangle = \alpha_{\pm}  \vert n, e \rangle + \beta_{\pm} \vert n+1, g \rangle, \label{eq:DressedStates}
\end{eqnarray}
with
\begin{eqnarray}
\frac{\alpha_{\pm}}{\beta_{\pm}} =  \frac{- \delta \pm \Omega }{2 \lambda}.
\end{eqnarray}
Note that $\vert 0,g \rangle$ is an eigenstate of the Hamiltonian with energy  $E_{-,0}=-\omega_{0}/2$.

\begin{figure}[h!]
\centering\includegraphics[scale=1]{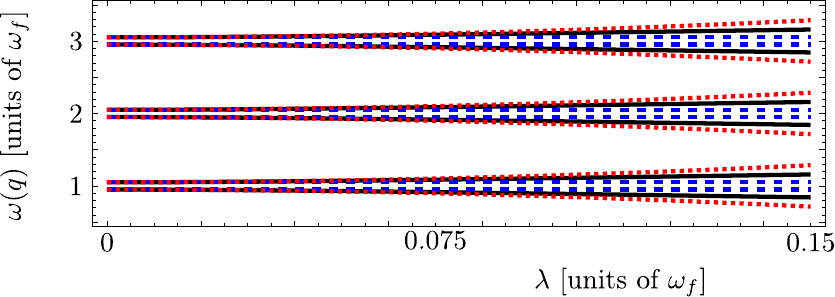}
\caption{(Color online) A segment of the dispersion relation in the weak-coupling regime. Exact closed form from rotating wave approximation (solid black), $0$th order perturbation (dashed blue) and second order perturbation (dotted red) are shown. We have used a detuning given by $\omega_{0}=1.1 ~\omega_{f}$.}
\label{fig:Fig2}
\end{figure}

In our photonic crystals, Eq. (\ref{eq:DiffSet}), we can approximate the dispersion relation, equivalent to the discrete spectrum found above, by proposing a collective proper mode and realizing that the three-term recurrence relations can be summarized by the tridiagonal matrix,
\begin{eqnarray}
H_{\pm,W} &=& H_{0}^{(\pm)} + P, \\
\left(H_{0}^{(\pm)}\right)_{i,j} &=& \left[  \omega_{f} j \mp \frac{\omega_{0}}{2} (-1)^{j} \right] \delta_{i,j}, \\
\left(P\right)_{i,j} &=& \lambda \left( \delta_{i,j+1} + \delta_{i+1,j} \right),
\end{eqnarray}
where the notation $\left( M \right)_{i,j}$ stands for the $(i,j)$th term of Matrix $M$ and the symbol $\delta_{a,b}$ is Kronecker's delta.
As $\lambda \ll \omega_{f}, \omega_{0}$, we can treat matrix $P$ as a perturbation on matrix $H_{0}$ and find the eigenvalues of $H_{\pm}$ up to second order corrections as the first order correction is equal to zero.
Thus, we obtain the approximated dispersion relation,
\begin{eqnarray}
\omega(q)^{(\pm)} &\approx&  \omega_{f} q  \mp \frac{ \omega_{0}}{2} (-1)^{q}   \left(1 + \frac{ 4 \lambda^{2}}{ \omega_{0}^2 - \omega_{f}^{2}} \right). \label{eq:DispRelWeak}
\end{eqnarray}
Figure \ref{fig:Fig2} shows good agreement between the dispersion relation given by the exact eigenvalues in the rotating wave approximation, Eq. (\ref{eq:EigValRWA}), and the perturbation approach, Eq. (\ref{eq:DispRelWeak}), at zero and second order.

%
\subsection{Strong-coupling regime: $\lambda \gg \omega_{f}, \omega_{0}$.}
In the case when the coupling parameter is larger than the field and transition frequencies, sometimes also called deep-coupling regime, it is possible to write the three term recurrence as
\begin{eqnarray}
H_{\pm,S} &=& H_{0} + P_{\pm}, \\
\left(H_{0}\right)_{i,j} &=& \lambda \left( \delta_{i,j+1} + \delta_{i+1,j} \right), \\
\left(P_{\pm}\right)_{i,j} &=&  \left[  \omega_{f} j \mp \frac{\omega_{0}}{2} (-1)^{j} \right] \delta_{i,j}.
\end{eqnarray}
Notice that lead order matrix is diagonalized via $H_{0} = V \Lambda V^{-1}$, where the diagonal matrix $\Lambda$ contains the values of the dispersion relation in its diagonal and the matrix $V$ has as columns the coefficients of the normal modes given by
\begin{eqnarray}
\left( V \right)_{j,q} = U_{j} \left( \frac{\mu(q)}{2 \lambda} \right), \quad \mu(q) \in \mathbb{R},
\end{eqnarray}
where the function $U_{n}(x)$ is the $n$th Chebyshev polynomial of the second kind evaluated at $x$;
i.e., the dispersion relation for this case is continuous. \\
The first order correction for the dispersion relation delivers,
\begin{eqnarray}
\omega(q)^{\pm} \approx \mu(q) + \int_{0}^{\infty} d\mu(q) \frac{\sum_{k=0}^{\infty} \left[U_{k}\left( \frac{\mu(q)}{2\lambda} \right) \right]^2 \left[ \omega_{f} k \mp \frac{\omega_{0}}{2} (-1)^{k} \right]}{ \sum_{j=0}^{\infty} \left[U_{j}\left( \frac{\mu(q)}{2\lambda} \right) \right]^2}.
\end{eqnarray}

%
\section{Collective modes}

For any given set of parameters, decomposition in normal modes delivers the three term recurrence mentioned before, 
\begin{eqnarray}
\left[a^{(\pm)}_{0} - \omega(q)\right] c^{(\pm,q)}_{0} + \lambda c^{(\pm,q)}_{1} &=&0,\label{eq:RecRel1} \\ 
\left[a^{(\pm)}_{j} - \omega(q)\right] c^{(\pm,q)}_{j} + \lambda ( c^{(\pm,q)}_{j-1} + c^{(\pm,q)}_{j+1}) &=&0, \label{eq:RecRel2}
\end{eqnarray}
with
\begin{eqnarray}
a^{(\pm)}_{j} &=& \omega_{f} j \mp \frac{\omega_{0}}{2}(-1)^{j},
\end{eqnarray}
where the coefficients $c^{(\pm,q)}_{k}$ are the $k$th coefficients of the $q$th collective mode corresponding to the proper value $\omega(q)$.
These coefficients are given by,
\begin{eqnarray}
c^{(\pm,q)}_{j} &=& \Pi_{k=0}^{j-1} s^{(\pm,q)}_{k} c^{(\pm,q)}_{0},
\end{eqnarray}
where we have used the continued fraction
\begin{eqnarray}
s^{(\pm,q)}_{j} &=& \frac{c^{(\pm,q)}_{j+1}}{c^{(\pm,q)}_{j}}, \\
&=& \frac{\lambda}{\omega(q) - a^{(\pm)}_{j+1} - \lambda s^{(\pm,q)}_{j+1}}, \\
&=& \frac{\lambda}{\omega(q) - a^{(\pm)}_{j+1} - \frac{\lambda^2}{\omega(q) - a^{(\pm)}_{j+2} - \ldots} },
\end{eqnarray}
where one can always set $c_{0}=1$ and normalize the semi-infinite set later.
In the weak-coupling case, the continued fraction is cut at the second term as the parameter $\lambda^2$ is negligible with respect to the field frequency.
This delivers a normal mode equivalent to that found in the rotating wave approximation treatment.

In any given regime, we can take the continued fraction result and use it to write the eigenvectors of the quantum Hamiltonian in the reduced form,
\begin{eqnarray}
| e_{\pm,m} \rangle &=&  \sum_{j=0}^{\infty} c_{j}^{(\pm,m)} |\pm,j\rangle,  \\
&=&  \tilde{c}_{0} \Pi_{k=0}^{\hat{n}-1} s_{k}^{(m)} \hat{n}!  e^{\hat{B}^{\dagger} - \hat{B}}  | \pm, 0 \rangle,
\end{eqnarray}
where $\tilde{c}_{0}$ is chosen such that $\langle e_{\pm,m} \vert e_{\pm,n} \rangle = \delta_{m,n}$.

For the case $\omega_{0}=0$, the amplitudes of the eigenvectors are well known and calculated by moving the differential set to Fourier domain and solving it there \cite{Peschel1998p1701}.
We want to mention here that it is trivial to obtain an equivalent form from the recurrence relations in Eqs.(\ref{eq:RecRel1}-\ref{eq:RecRel2}) by setting $\omega_{0}=0$, $c_{j}^{(+,m)}=c_{j}^{(-,m)}= c_{j}(\omega_{q})\equiv c_{j}$, $\omega(q)\equiv \omega$, and $c_{0}=1$,
\begin{eqnarray}
c_{j}&=& \frac{\pi}{\omega_{f}} \left\{ Y_{j-\frac{\omega}{\omega_{f}}}\left( -\frac{2 \lambda}{\omega_{f}}\right) \left[ (\omega - j \omega_{f}) J_{-\frac{\omega}{\omega_{f}}}\left( -\frac{2 \lambda}{\omega_{f}}\right) - \lambda  J_{1-\frac{\omega}{\omega_{f}}}\left( -\frac{2 \lambda}{\omega_{f}}\right) \right] \right. + \nonumber \\
&& \left. - J_{j-\frac{\omega}{\omega_{f}}}\left( -\frac{2 \lambda}{\omega_{f}}\right) \left[ (\omega - j \omega_{f}) Y_{-\frac{\omega}{\omega_{f}}}\left( -\frac{2 \lambda}{\omega_{f}}\right) - \lambda  Y_{1-\frac{\omega}{\omega_{f}}}\left( -\frac{2 \lambda}{\omega_{f}}\right)\right]  \right\},
\end{eqnarray}
where the symbols $J_{\alpha}(x)$ and $Y_{\alpha}(x)$ stand for the modified Bessel functions of the first and second kind, and the coefficients are given up to a normalization factor.

%
\section{Propagation examples}

\begin{figure}[h!]
\centering\includegraphics[scale=1]{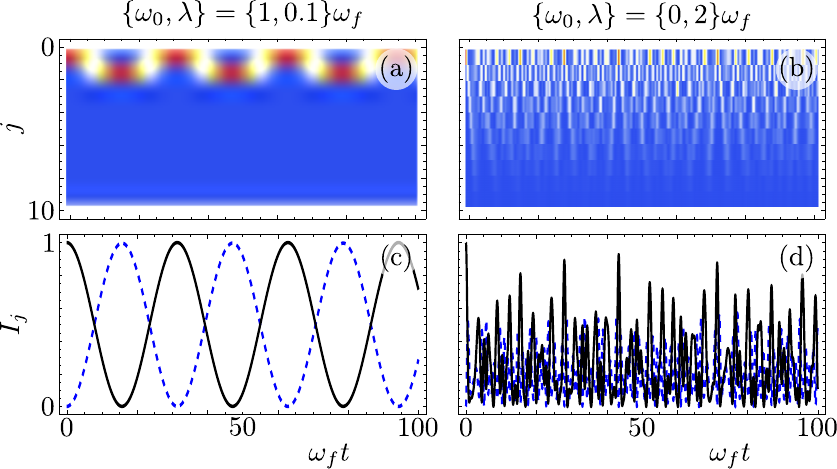}
\caption{(Color online) Examples of propagation in our negative parity photonic crystal. The left column (a,c) shows the case of weak-coupling, $\lambda = 0.1 ~\omega_{f}$, and the right (b,d) the case of strong-coupling, $\lambda = 2 ~ \omega_{f}$.
The first row (a,b) depicts intensity propagation on the first ten waveguides of a total of five thousand when light impinges the zeroth waveguide.
The second row (c,d) shows the normalized intensity at the $0$th (solid black) and first (dashed blue) waveguides.
The dimensionless time parameter $\omega_{f} t$ is equivalent to the typical dimensionless propagation parameter.}
\label{fig:Fig3}
\end{figure}

The dressed state basis that diagonalize our model in the weak-coupling regime, Eq.(\ref{eq:DressedStates}), implies that starting in a state of the kind $\vert j, e \rangle$ or $\vert j+1, g \rangle$, with $j\ge0$, will produce coherent oscillations.
Such an initial state translates to laser light impinging the $j$th or $(j+1)$th waveguide of one of the lattices; the case $j$ even (odd) corresponds to the crystal  $H_-$ ($H_+$ ).
Figure \ref{fig:Fig3}(a) shows the propagation of light impinging at the $0$th waveguide of our photononic crystal $H_{-}$ in the weak-coupling regime which is equivalent to an initial state $\vert 0,e \rangle$ in the quantum optics model.
Accordingly, we observe the intensity oscillate between the $0$th and first wave corresponding to an oscillation between the  $\vert 0,e \rangle$ and $\vert 1,g \rangle$ states in the quantum optics model.
Figure \ref{fig:Fig3}(c) presents a detail of the intensity at the $0$th and first waveguide.
This normalized intensity at the 0th waveguide is proportional to the probability of finding the time evolution of the quantum system back in the initial state, $I_{0} \leftrightarrow P_{-,0}$, with 
\begin{eqnarray}
P_{-,0} = \vert \langle -,0 \vert \psi(t) \rangle \vert^2, \quad \vert \psi(0) \rangle = \vert -,0 \rangle. 
\end{eqnarray}

In quantum optics literature, it is known that the quantum Rabi model produces coherent oscillations in the two-level system inversion for $\omega_{0}=0$~\cite{Casanova2010p263603}.
In our model, due to the continuous spectra, all proper states are scattering states and, in the general case, at most we observe partial recovery of the original state when the field starts localized at a given waveguide.
Figure \ref{fig:Fig3}(b) shows the propagation of light impinging at the $0$th waveguide of the photonic lattice $H_{+}$ in the strong-coupling regime, which is equivalent to an initial state $\vert 0,g \rangle$ in the quantum optics model; Figure \ref{fig:Fig3}(d) focus on the intensity at the first two waveguides.
Again, the normalized intensity at the 0th waveguide is proportional to the probability of finding the time evolution of the quantum system back in the initial state, $I_{0} \leftrightarrow P_{+,0} = \vert \langle +,0 \vert \psi(t) \rangle \vert^2$, with $\vert \psi(0) \rangle = \vert +,0 \rangle$, for this case.
Thus, our model presents both similar and different behaviors from the full quantum Rabi model in the weak- and strong-coupling regimes, in that order.

We can go further than comparing with the full quantum Rabi problem. 
For $\omega_{0}=0$, it is known that single waveguide initial states far from the edge will produce a breather mode evolution and will reconstruct periodically, while multiple-waveguide initial states with Gaussian weight distributions will present Bloch oscillations~\cite{Peschel1998p1701,Pertsch1999p4752}.
Also, if we consider a double contiguous waveguide input with a phase difference between the weights of the components, 
\begin{eqnarray}
\vert \psi_{k}(0) \rangle = \frac{1}{\sqrt{2}} \left( \vert k \rangle + e^{i \phi} \vert k+1 \rangle \right), \quad k \gg 0, \label{eq:Superposition}
\end{eqnarray}
it is possible to obtain a periodical reconstruction~\cite{Thompson2011p214302}.
This is witnessed by the fidelity, 
\begin{eqnarray}
\mathcal{F} = \vert \langle \psi(0) \vert \psi(t) \rangle\vert^2. \label{eq:Fidelity}
\end{eqnarray} 
Such an initial state produces a break in the symmetry of the breather mode of a single input for $\phi \ne 0$~~\cite{Thompson2011p214302}.  
This breaking in the propagation symmetry is accompanied by a variable center of mass of the beam,
\begin{eqnarray}
x_{cm} &=&  \langle \psi(t)  \vert \hat{n} \vert \psi(t) \rangle,  \label{eq:CenterMass} \\
&=& \sum_{j}^{\infty} j \vert c_{j} \vert ^{2}, 
\end{eqnarray} 
where the amplitudes $c_{j}$ are normalized.

\begin{figure}[h!]
\centering\includegraphics[scale=1]{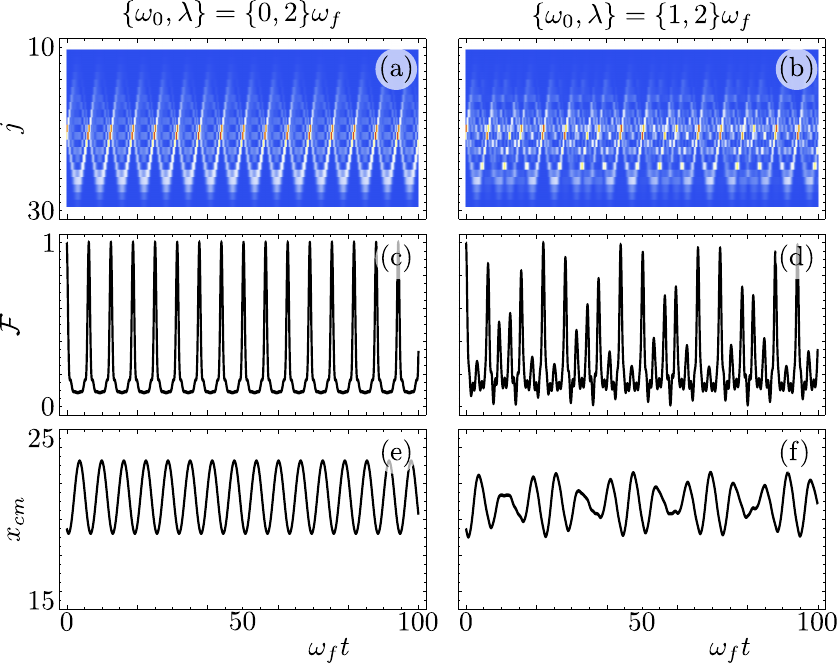}
\caption{(Color online) Examples of propagation in our positive parity photonic crystal in the strong-coupling regime with $\lambda = 2 ~\omega_{f}$.  The initial state is the balanced superposition of light impinging the 20th and 21th waveguide with a phase difference $\phi = \pi/6$.
The left column (a,c,e) shows the case, $\omega_{0} = 0$, and the right column (b,d,f) the case, $\omega_{0} = \omega_{f}$.
The first row (a,b) depicts intensity propagation on the waveguides from 10th to 30th of a total of five thousand.
The second row (c,d) shows the evolution of the Fidelity, Eq.(\ref{eq:Fidelity}).
The third row (e,f) shows the evolution of the center of mass of the beam, Eq.(\ref{eq:CenterMass}).
The dimensionless time parameter $\omega_{f} t$ is equivalent to the typical dimensionless propagation parameter.}
\label{fig:Fig4}
\end{figure}

Figure 4 shows the propagation of double waveguide initial state, $\vert \psi_{k}(0)\rangle$ in Eq.(\ref{eq:Superposition}) with $k=20$ and $\phi = \pi/6$.
The left column shows the case where $\omega_{0}=0$; it is possible to see that a nonsymmetric breather mode is formed upon propagation, Fig.\ref{fig:Fig4}(a), with periodic reconstruction, Fig.\ref{fig:Fig4}(c), and a so-called ratchet like behavior of the center of mass of the beam, Fig.\ref{fig:Fig4}(e).
The introduction of the binary modification, $\omega_{0}=\omega_{f}$, destroys the nonsymmetric propagation breather mode, Fig.\ref{fig:Fig4}(b), with partial reconstruction of the input, Fig.\ref{fig:Fig4}(d), and a quasi-periodical behavior of the center of mass of the beam, Fig.\ref{fig:Fig4}(f).

All numerical simulations correspond to a finite photonic lattice of size 5000.

%
\section{Conclusion}

We have proposed a set of two photonic crystals that classically simulates a new radiation-matter interaction where a two-level system is driven by just the phase of a quantum field.
Up to our knowledge such a radiation--matter interaction systems does not occur in nature and has not been proposed before.

We show that it is possible to determine exactly the dispersion relation of the photonic waveguide lattices in the so-called weak-coupling regime and that in the strong-coupling regime we can use perturbation theory to approximate the dispersion relation up to second order perturbation.
In the first case, the dispersion relation is discrete and, in the latter, continuous.
Accordingly, the spectra of the radiation--matter interaction model is equivalent to that of the classical simulator.

The normal modes of the crystals are easily expressed in terms of continued fractions as a function of the dispersion relation.
In a simplified version, they can be expressed in terms of modified Bessel functions of the first and second kind.

In the simplified version, an initial state consisting of light impinging two contiguous waveguides with a phase difference between them produces a phase controlled, so-called ratchet-like behavior of the center of mass of the beam upon propagation. 
This behavior is attenuated and becomes noisy in the most general model as the extra parameter increases in value. 

Our optical realization of a phase driven two-level system provides a scheme to explore an interesting process that is not accessible by usual means; i.e. cavity-QED or trapped ions.
And describes a phase controlled phenomenon that may be realized in other radiation--matter interaction systems. 

\section*{Acknowledgement}
The authors are grateful to an anonymous reviewer for his valuable comments and bringing forward important references. 

%
\bibliographystyle{model1-num-names}

\end{document}